\documentclass[final,3p,times]{elsarticle}

\usepackage{graphicx}

\usepackage{amssymb}

\usepackage{lineno}
  
\newcounter{bla}

\journal{Computer Physics Communications}

\usepackage{amsmath}
\usepackage[usenames,dvipsnames]{xcolor}
\usepackage{tikz}
\usetikzlibrary{positioning}
\usepackage{beramono}
\usepackage[T1]{fontenc}
\usepackage{listings}
\lstdefinelanguage{Julia}%
  {morekeywords={abstract,break,case,catch,const,continue,do,else,elseif,%
      end,export,false,for,function,immutable,import,importall,if,in,%
      macro,module,otherwise,quote,return,switch,true,try,type,typealias,%
      using,while,begin},%
   sensitive=true,%
   morecomment=[l]\#,%
   morecomment=[n]{\#=}{=\#},%
   morestring=[s]{"}{"},%
   morestring=[m]{'}{'},%
}[keywords,comments,strings]%

\usepackage{courier}
\usepackage{caption}
\usepackage{subcaption}
\usepackage[hidelinks]{hyperref}

\newcommand{\eg}{e.g.}
\newcommand{\ie}{i.e.}

\def\bm#1{\mbox{\boldmath{$#1$}}}

\begin{document}

\begin{frontmatter}



\title{Biofilm.jl: a fast solver for one-dimensional biofilm chemistry and ecology}


\author[a]{Mark Owkes\corref{author}}
\author[a]{Kai Coblentz}
\author[a]{Austen Eriksson}
\author[a]{Takumi Kammerzell}
\author[b]{Philip S. Stewart}

\cortext[author] {Corresponding author.\\\textit{E-mail address:} mark.owkes@montana.edu}
\address[a]{Mechanical \& Industrial Engineering, Montana State University}
\address[b]{Chemical \& Biological Engineering, Montana State University}

\begin{abstract}






Biofilms are communities of microorganisms that grow on virtually all surfaces with sufficient nutrients including aquatic and industrial water systems and medical devices.  Biofilms are complex, structured communities where the interplay of growth, metabolism, and competition between species interact with physical processes of diffusion, convection, attachment, and detachment.  This work describes a model of a one-dimensional biofilm in a stirred tank reactor that incorporates these complexities.  The model is implemented in the modern Julia programming language providing an efficient tool for studying a large variety of biofilms and the intricate communities the microorganisms create.  Details of the new software, known as Biofilm.jl, including the mathematical model and organization and execution of the code, are provided.  Examples of biofilms modeled using Biofilm.jl are presented such as a single heterotroph, a sulfide-oxidizing bacteria (SOB) and sulfate-reducing bacteria (SRB), and a phototroph. Postprocessing tools are described that allow a Biofilm.jl user to make plots and extract specific values from the solution and explore the simulated biofilm results. 

\end{abstract}


\end{frontmatter}



{\bf PROGRAM SUMMARY}

\begin{small}
\noindent
{\em Program Title:} Biofilm.jl                                          \\
{\em CPC Library link to program files:} (to be added by Technical Editor) \\
{\em Developer's repository link:} https://github.com/markowkes/Biofilm.jl \\
{\em Code Ocean capsule:} (to be added by Technical Editor)\\
{\em Licensing provisions(please choose one):} MIT  \\
{\em Programming language:} Julia (Developed on v1.8, tested on v1.7)  \\
{\em Supplementary material:}  https://markowkes.github.io/Biofilm.jl  \\
{\em Nature of problem:}\\ 
  This software solves for the temporal and spatial dynamics of an arbitrary number of substrates (or products) and particulates (biomass species) in a one-dimensional biofilm.  The model includes the growth of particulates and the associated substrate consumption (or production), the biofilm thickness dynamics due to growth within and detachment from the top of the biofilm.  Additionally, source terms can be used to model the death of biomass or other effects.  Discontinuous inputs, such as the diurnal cycle or periodic dosing, can be included.\\
{\em Solution method:}\\ 
  The software solves for the temporal dynamics of particulates, substrates, and biofilm thickness, which are described by differential equations.  These equations are discretized using a finite volume method and organized into a single system of ordinary differential equations that are solved using the DifferentialEquations.jl library.  The software includes a collection of postprocessing tools to assist the user with exploring the simulation results.


\end{small}

\section{Introduction}
\label{sec:intro}

Biofilms are communities of microorganisms that are attached to a surface~\cite{otoole_biofilm_2000} and predominate in almost all wet environments with sufficient nutrients~\cite{costerton_microbial_1995}.  Biofilms can form on a variety of surfaces including, for example, natural aquatic systems, teeth (dental plaque), and water system piping~\cite{percival_introduction_2011}.  Furthermore, biofilms are not confined to solid-liquid surfaces and can be found at solid-air or liquid-liquid interfaces~\cite{percival_introduction_2011}.  Biofilms have been shown to be associated with many human diseases and grow on a variety of medical devices and implanted biomaterials~\cite{donlan_biofilms_2002,douglas_candida_2003,ramage_candida_2006}. 

Biofilm microorganisms, known as particulates, grow and consume or produce substrates. For example, an aerobe is a particulate that will consume the substrate oxygen when it grows.  A biofilm may consist of a single particulate or many particulates.  When there are many particulates, interactions with consumed or produced substrates and other physical phenomena can lead to highly structured multispecies communities~\cite{stoodley_biofilms_2002}.  Interestingly, these multispecies biofilms have been demonstrated to be profoundly different from planktonic (drifting or floating) cells with  increased growth rates, enhanced tolerance to antibacterial agents, and changes in gene expression~\cite{percival_introduction_2011,costerton_microbial_1995}.

Mathematical models have played an important role in the biofilm field because the complexity of these systems - in which biological phenomena such as growth, metabolism, and competition between species interact with physical phenomena such as diffusion, convection, attachment and detachment -  leads to non-intuitive emergent properties. These behaviors can be accessed rigorously through computational approaches. Many different model structures from one-dimensional flat slabs to fully three-dimensional particle-based simulations have been derived \cite{wimpenny_unifying_1997,lardon_idynomics_2011,kreft_individual-based_2001,picioreanu_particle-based_2004,xavier_general_2005,alpkvist_multidimensional_2007,chambless_three-dimensional_2006,chambless_three-dimensional_2007} 
. Several comprehensive reviews describe and discuss the variety of models in detail 
\cite{chaudhry_review_1998,wang_review_2010,horn_modeling_2014}.

One of the earliest model structures was a one-dimensional version in which the biofilm is treated as a uniformly thick flat slab 
\cite{wanner_multispecies_1986,wanner_mathematical_1996}.
Solute transport into and out of the film is assumed to be governed by Fick’s Law. The competition for space within the biofilm by different microbial species and abiotic particles is analyzed by differential material balances on each particulate. The dynamics of biofilm thickness are controlled by the net balance of integrated growth and detachment from the top layer of the biofilm.

Though relatively simplistic, these types of models successfully capture such phenomena as: coexistence of aerobes and anaerobes in close proximity within the same biofilm and stratification of these two types of microorganisms 
\cite{wanner_multispecies_1986,arcangeli_modelling_1999}, activity of biofilm-based wastewater treatment processes 
\cite{wanner_mathematical_1996,arcangeli_modelling_1999,wanner_biofilm_2004}, 
slow growth and dormancy in some regions of the biofilm 
\cite{stewart_biofilm_1994,roberts_modeling_2004,roberts_modelling_2005}, 
reduced antimicrobial susceptibility in the biofilm mode of growth 
\cite{stewart_biofilm_1994,stewart_modeling_1996,roberts_modeling_2004,roberts_modelling_2005}, 
and plasmid transfer 
\cite{beaudoin_mobilization_1998}
.

Here we recapitulate the classic 1D biofilm model with an updated code and solver. This has the potential to be used as an educational and research tool, replacing outdated code developed at the Montana State University Center for Biofilm Engineering. This legacy code, now three decades old, is relatively slow and is not easily modified to include custom kinetic expressions or time-dependent boundary conditions.  The updated code, known as Biofilm.jl, is written in the modern Julia programming language \cite{bezanson_julia_2017} which is designed to be fast while being a relatively easy language to write, run, and process results.  Within Julia, Biofilm.jl leverages the state-of-the-art differential equation solvers~\cite{rackauckas_differentialequationsjl_2017} that help Biofilm.jl achieve fast and stable solutions to many biofilm problems.

The manuscript provides a detailed overview of the biofilm model and governing equations in Section~\ref{sec:GovEqs} and the numerical methods and implementation are presented in Section~\ref{sec:num_method}.  Specifics on how the code is organized and can be executed are in Section~\ref{sec:org_exc}.  Finally, results on applying the code to five different test cases are provided in Section~\ref{sec:examples}.

\section{Governing Equations}
\label{sec:GovEqs}
Biofilm.jl simulates a one-dimensional biofilm within a stirred tank reactor.  More specifically, the concentrations of particulates $\bm{X}_{t}$ and substrates $\bm{S}_t$ within the tank are solved along with the particulate volume fractions $\bm{P}_{b}$ and substrate concentrations $\bm{S}_{b}$ within the biofilm.  The thickness of the biofilm $L_f$ is also computed.  The code is written to solve for an arbitrary number of particulates and substrates, $N_x$ and $N_s$, respectively. 


\subsection{Tank Equations}
The temporal dynamics of particulate concentrations in the tank environment are described by
\begin{equation}
\label{eq:dXtdt}
\frac{d X_{t:j}}{dt} = \mu_j(\bm{S}_t) X_{t:j} - \frac{Q X_{t:j}}{V} + \frac{v_\mathrm{det} A X_{b:j}(L_f)}{V} + \mathrm{src}_{X:j}
\end{equation}
for $j=1,\dots,N_x$.  The terms on the right-hand-side (RHS) are 
1) the growth of the particulate in the tank, 
2) transport due to flow out of the tank, 
3) transfer to the tank due to the detachment of particulates from the biofilm, and
4) a source term of particulates.
The variables in the equation are as follows, $t$ is time, 
$\mu_j(\bm{S}_t)$ is the growth rate of the $j^\mathrm{th}$ particulate and is dependent on the substrate concentrations within the tank $\bm{S}_t$, 
$Q$ is the flowrate into and out of the tank, 
$V$ is the volume of the tank, 
$A$ is the area of the biofilm, 
$X_{b:j}(L_f) = \rho_j P_{b:j}(L_f)$ is the $j^\mathrm{th}$ particulate concentration at the top of the biofilm that can be computed from the particulate density $\rho_j$ and volume fraction at the top of the biofilm $P_{b:j}(L_f)$, and
$\mathrm{src}_{X:j}$ is the source term for the $j^\mathrm{th}$ particulate. 
The detachment velocity $v_\mathrm{det}$ is modeled using
\begin{equation}
    \label{eq:vdet}
    v_\mathrm{det}\equiv K_\mathrm{det} L_f^2.
\end{equation}

The substrate concentrations in the tank environment are described by
\begin{equation}
\label{eq:dStdt}
\frac{d S_{t:k}}{dt} = -\sum_{j=1}^{N_x} \frac{\mu_j(\bm{S}_t) X_{t:j}}{Y_{j,k}} + \frac{Q S_{\mathrm{in}:k}}{V} - \frac{Q S_{t:k}}{V} - \frac{A S_             {t:k}^\mathrm{flux}} {V} + \mathrm{src}_{S:k}
\end{equation}
for $k=1,\dots,N_s$, where $Y_{j,k}$ is the biomass yield coefficient on the substrate, 
$S_{\mathrm{in}:k}$ is the influent substrate concentration,
$S_{t:k}^\mathrm{flux}$ is the flux of substrates into the biofilm from the tank due to diffusion, and 
$\mathrm{src}_{S:k}$ is the source term for the $k^\mathrm{th}$ substrate. 
The terms on the RHS are 
1) consumption of substrates due to the growth of the particulate in the tank, 
2) transport due to flow into the tank, 
3) transport due to flow out of the tank,
4) transfer of substrates into the biofilm due to diffusion, and
5) a source term of substrates.
   
\subsection{Biofilm Equations}
The particulates within the biofilm are described using their volume fractions $P_{b}$ and are described with
\begin{equation}
\label{eq:dPbdt}
\frac{\partial P_{b:j}}{\partial t} = 
\mu_j(\bm{S}_b) P_{b:j} 
- \frac{\partial v(z) P_{b:j}}{\partial z} 
+ \frac{\mathrm{src}_{X:j}}{\rho_j}
\end{equation}
for $j=1,\dots,N_x$ and 
$P_{b:j}(t,z)$ is the $j^\mathrm{th}$ particulate within the biofilm and varies with time and location within the biofilm. 
The terms on the RHS are 
1) the growth of the particulate in the biofilm, 
2) transport through the biofilm due to the growth velocity $v_i$, and 
3) source term of particulate at $i^\mathrm{th}$ location in biofilm.

The growth velocity $v(z)$ is the vertical velocity of the biofilm due to the growth and sources of particulates deeper within the biofilm.  In general, this velocity can be written as an integral that captures growth and sources from the bottom of the biofilm ($z=0$) to some location $z$ within the biofilm, \ie,  
\begin{equation}
\label{eq:growthvelocity}
v(z)=  \int_{z'=0}^{z}{\sum_{j=1}^{N_x} \frac{1}{P_\mathrm{tot}}\left(\mu_j(\bm{S}_{b}(z')) P_{b:j} + \frac{\mathrm{src}_{X:j}}{\rho_j}\right) ~dz'}
\end{equation}
where $P_\mathrm{tot}=\sum_{j=1}^{N_x}{P_{b:j}}$
   
The dynamics of substrates within the biofilm are described by 
\begin{equation}
\label{eq:dSbdt}
\frac{\partial S_{b:k}}{\partial t} = 
D_{b:k}\frac{\partial^2 S_{b:k}}{\partial z^2} 
- \sum_{j=1}^{N_x} \frac{\mu_j(\bm{S}_{b}) X_{b:j}}{Y_{j,k}}
+ \mathrm{src}_{S:k}
\end{equation}
for $k=1,\dots,N_s$, where $D_{b:k}$ is the effective diffusion coefficient of the $k^\mathrm{th}$ substrate within the biofilm.
The terms on the RHS are 
1) diffusion of substrates in the biofilm,
2) consumption of substrates due to the growth of the particulate in the biofilm, and
3) source term of substrates.  Note that the top boundary condition of the diffusion term must match the flux of substrate into the biofilm $S_{t:k}^\mathrm{flux}$ and provides a connection between the tank and biofilm dynamics. 

\subsubsection{Biofilm Thickness}
The thickness of the biofilm $L_f$ is described by 
\begin{equation}
\label{eq:dLfdt}
\frac{d L_f}{dt} = v(L_f) - v_\mathrm{det}
\end{equation}
where the first term on the RHS is the growth velocity at the top of the biofilm (Eq.~\ref{eq:growthvelocity}) and the second term is the detachment velocity (Eq.~\ref{eq:vdet}).

\section{Discretization and Numerical Methods}\label{sec:num_method}
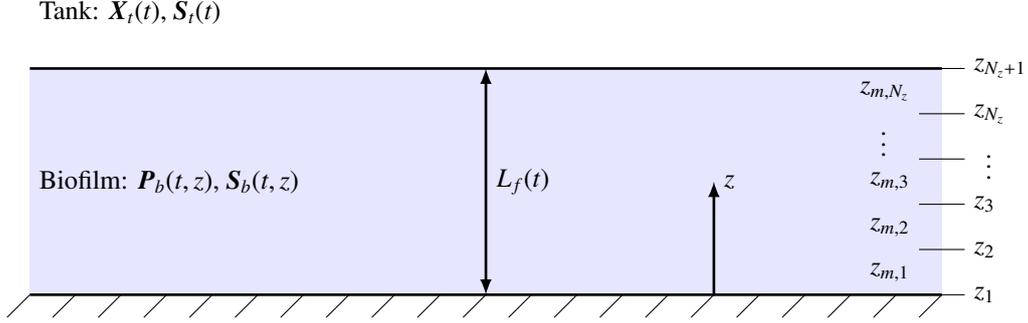
\begin{figure}[htbp]
    \centering
    \begin{tikzpicture}[scale=3]
        \fill[blue,opacity=0.1] (0,0) rectangle (4,1);
        \draw[line width=1] (0,0) -- (4,0);
        \foreach \x in {0.0,0.2,...,4.001}{
            \draw[] (\x,0) -- +(-0.1,-0.1);
        }
        \draw[line width=1] (0,1) -- (4,1);
        \node[anchor=west] (A) at (0.0,1.25) {Tank: $\bm{X}_t(t)$, $\bm{S}_t(t)$};
        \node[anchor=west] (B) at (0.0,0.5) {Biofilm: $\bm{P}_b(t,z)$, $\bm{S}_b(t,z)$};
        \draw[line width=1,latex-latex] (2.0,0) -- node[pos=0.5,right]{$L_f(t)$} +(0,1);
        \draw[line width=1,-latex] (3.0,0) -- node[pos=1.0,right]{$z$} +(0,0.5);
        \foreach \y in {0,0.2,...,1.001}{
            \draw[] (3.9,\y)  -- +(0.2,0.0);
        }
        \node[anchor=east] at (3.9,0.1) {$z_{m,1}$};
        \node[anchor=east] at (3.9,0.3) {$z_{m,2}$};
        \node[anchor=east] at (3.9,0.5) {$z_{m,3}$};
        \node[anchor=east] at (3.8,0.7) {$\vdots$};
        \node[anchor=east] at (3.9,0.9) {$z_{m,N_z}$};
        \node[anchor=west] at (4.1,0.0) {$z_{1}$};
        \node[anchor=west] at (4.1,0.2) {$z_{2}$};
        \node[anchor=west] at (4.1,0.4) {$z_{3}$};
        \node[anchor=west] at (4.15,0.6) {$\vdots$};
        \node[anchor=west] at (4.1,0.8) {$z_{N_z}$};
        \node[anchor=west] at (4.1,1.0) {$z_{N_z+1}$};
    \end{tikzpicture}
    \caption{Geometry used in the model.  The tank is a stirred tank reactor with particulate $\bm{X}_t(t)$ and substrate $\bm{S}_t(t)$ concentrations. The one-dimensional biofilm contains particulate volume fractions $\bm{P}_b(t,z)$ and substrate concentrations $\bm{S}_b(t,z)$ that vary with $z$.  The thickness of the biofilm is $L_f(t)$. The biofilm is discretized with $N_z$ grid cells in the $z$-direction and the middle and faces of the grid cells are shown with $z_m$ and $z$, respectively.}
    \label{fig:geometry}
\end{figure}
The model solves for the particulate and substrate concentrations within the tank and the particulate volume fractions and substrate concentrations in the biofilm along with the biofilm thickness.  The tank is assumed to be a stirred tank reactor and thus the tank concentrations are only a function of time.  The variations within the biofilm are, however, important and a one-dimensional grid is used with $N_z$ grid cells.  The variables are stored at the cell centers denoted $z_{m,i}$ for $i = 1,\dots,N_z$.  Cell faces are denoted $z_i$ for $i = 1,\dots,N_z+1$.

The governing equations consist of five time-dependent differential equations (Eqs.~\ref{eq:dXtdt}, \ref{eq:dStdt}, \ref{eq:dPbdt}, \ref{eq:dSbdt}, and \ref{eq:dLfdt}) that are solved at the same time using built-in ODE solvers in Julia's DifferentialEquations.jl library~\cite{rackauckas_differentialequationsjl_2017}.  To use the solvers the dependent variables are organized in a single vector such that the system of equations can be written as 
\begin{equation}
    \label{eq:singleODE}
    \frac{d \mathrm{\mathbf{sol}}}{dt}=\mathrm{\mathbf{rhs}}
\end{equation}
where 
\begin{equation*}
    \mathrm{\mathbf{sol}}=[\bm{X}_t,\bm{S}_t,\bm{P}_b,\bm{S}_b,Lf]^\mathsf{T}
\end{equation*}
where $[\cdot]^\mathsf{T}$ indicates transpose.  The bold terms on the RHS of the previous equation contain the different particulates and substrates and can be written as
\begin{align*}
    \bm{X} &= [X_{t:1},X_{t:2},\dots,X_{t:N_x}]^\mathsf{T},\\
    \bm{S} &= [S_{t:1},S_{t:2},\dots,S_{t:N_s}]^\mathsf{T},\\
    \bm{P}_b &= [\bm{P}_{b:1},\bm{P}_{b:2},\dots,\bm{P}_{b:N_x}]^\mathsf{T},\text{ and}\\
    \bm{S}_b &= [\bm{S}_{b:1},\bm{S}_{b:2},\dots,\bm{S}_{b:N_s}]^\mathsf{T}.
\end{align*}
These last two expressions contain the particulate and substrates within the biofilm.  The biofilm is discretized with $N_z$ grid points and therefore each of these variables should be written as a vector of values at each grid location, \ie, 
\begin{align*}
    \bm{P}_{b:j}=[P_{b:j,1},P_{b:j,2},\dots,P_{b:j,N_z}]^\mathsf{T}\\
    \bm{S}_{b:j}=[S_{b:j,1},S_{b:j,2},\dots,S_{b:j,N_z}]^\mathsf{T}
\end{align*}
All together $\mathrm{\mathbf{sol}}$ is a vector with 
\begin{equation}
    \label{eq:Nvar}
    N_\mathrm{var} = N_x + N_s + N_x\cdot N_z + N_s\cdot N_z + 1
\end{equation}
variables.  

Similarly, $\mathrm{\mathbf{rhs}}$ is a vector of length $N_\mathrm{var}$ defined with 
\begin{equation*}
    \mathrm{\mathbf{rhs}}=\left[\frac{d \bm{X}}{dt},\frac{d \bm{S}}{dt},\frac{d \bm{P}_b}{dt},\frac{d \bm{S}_b}{dt},\frac{d Lf}{dt}\right]^\mathsf{T}
\end{equation*}
where the right-hand sides are defined using the equations provided in Section~\ref{sec:GovEqs}.  The RHS of the ordinary differential equations, Eq.~\ref{eq:dXtdt}, \ref{eq:dStdt}, and \ref{eq:dLfdt}, can be used as written.  However, the partial differential equations (Eq.~\ref{eq:dPbdt} and \ref{eq:dSbdt}) that describe the particulate volume fractions and substrate concentrations within the biofilm need to be discretized before they can be used with Julia's ODE solvers.  Biofilm.jl uses a finite volume discretization to approximate the spatial derivatives.  

To discretize Eq.~\ref{eq:dPbdt} for the particulate volume fractions in the biofilm, the equation is written at each grid cell, \ie,
\begin{equation}
\label{eq:dPbdt_discrete}
\frac{\partial P_{b:j,i}}{\partial t} = 
\mu_j(\bm{S}_{b:i}) P_{b:j,i} 
- \frac{\partial v_i P_{b:j,i}}{\partial z} 
+ \frac{\mathrm{src}_{X:j,i}}{\rho_j}
\end{equation}
for $j=1,\dots,N_x$ and $i=1,\dots,N_z$.
The spatial derivative is discretized as 
\begin{equation*}
    \frac{\partial v_i P_{b:j,i}}{\partial z} = \frac{P_{b:j,i+1}^\mathrm{flux} - P_{b:j,i}^\mathrm{flux}}{\Delta z}
\end{equation*}
where $\Delta z$ is size of the grid cell and the flux on the $i^\mathrm{th}$ face ($z_i$ location) is defined as
\begin{equation*}
    P_{b:j,i}^\mathrm{flux} = v(z_i) P_{b:j,i-1}.
\end{equation*}
for $j=1,\dots,N_x$ and $i=2,\dots,Nz+1$.
This expression approximates the volume fraction at the cell face using the volume fraction in the cell below the face introducing upwinding that stabilizes the solution.  Note that a zero-flux condition is used at the bottom of the biofilm, $P_{b:j,1}^\mathrm{flux}=0$.  In the previous equation $v(z_i)$ is computed using the discrete form of Eq.~\ref{eq:growthvelocity} which is

\begin{equation}
\label{eq:growthvelocity_discrete}
v(z_i)=  \sum_{n=1}^{i-1} \sum_{j=1}^{N_x} \frac{1}{P_\mathrm{tot}}\left(\mu_j(\bm{S}_{b:i}) P_{b:j,i} + \frac{\mathrm{src}_{X:j,i}}{\rho_j}\right) ~\Delta z
\end{equation}
and includes contributions from growth and sources in all cells below the $i^\mathrm{th}$ face. 

Similarly, to discretize Eq.~\ref{eq:dSbdt} for the substrate concentrations in the biofilm, the equation is written at each grid cell, \ie,
\begin{equation}
\label{eq:dSbdt_discrete}
\frac{\partial S_{b:k,i}}{\partial t} = 
D_{b:k}\frac{\partial^2 S_{b:k,i}}{\partial z^2} 
- \sum_{j=1}^{N_x} \frac{\mu_j(\bm{S}_{b:i}) X_{b:j,i}}{Y_{j,k}}
+ \mathrm{src}_{S:k,i}
\end{equation}
for $k=1,\dots,N_s$ and $i=1,\dots,N_z$.  
The diffusion term with a second-order derivative is discretized using 
\begin{equation*}
    D_{b:k}\frac{\partial^2 S_{b:k,i}}{\partial z^2} 
    = \frac{S_{b:k,i+1}^\mathrm{flux} - S_{b:k,i}^\mathrm{flux}}{\Delta z}
\end{equation*}
and the flux on the $i^\mathrm{th}$ face ($z_i$ location) is defined using a second-order, central finite difference operator that can be written as
\begin{equation}
    S_{b:k,i}^\mathrm{flux} = D_{b:k}\frac{S_{b:k,i} - S_{b:k,i-1}}{\Delta z}
\end{equation}
for $k=1,\dots,N_s$ and $i=2,\dots,N_z$
The flux requires boundary conditions at the top and bottom of the biofilm.  A zero-flux condition is used at the bottom of the biofilm, \ie, $S_{b:k,1}^\mathrm{flux}=0$.   For the boundary condition at the top of the biofilm, the diffusive flux from the tank is matched with the diffusive flux at the top of the biofilm, i.e.,
\begin{align}
S_{t:k}^\mathrm{flux} &= S_{b:k,N_z+1}^\mathrm{flux} \nonumber\\
D_{t:k}\frac{d S_{t:k}}{dz} &= D_{b:k}\frac{d S_{b:k,N_z}}{dz} \label{eq:fluxMatch}
\end{align}
where $D_\mathrm{t}$ is the diffusion coefficient in the tank.  A boundary layer within the tank at the top of the biofilm of thickness $L_L$ is introduced to evaluate the derivative in the tank.  Discretizing these derivatives requires introducing the substrate concentration at the top of the biofilm $S_\mathrm{top}$, which when combined with finite differences, leads to
\begin{equation*}
D_{t:k}\frac{S_{t:k} - S_\mathrm{top}}{L_L} = D_{b:k}\frac{S_\mathrm{top} - S_{b:k,N_z}}{\Delta z/2}.
\end{equation*}
Rearranging provides an expression for $S_\mathrm{top}$, \ie,
\begin{equation}
    \label{eq:Stop}
    S_\mathrm{top} = \frac{ D_{t:k} (\Delta z/2) S_{t:k}    + D_{b:k} L_L S_{b:k,N_z}  }
    { D_{t:k} (\Delta z/2) + D_{b:k} L_L }.
\end{equation}
With $S_\mathrm{top}$ defined, either the LHS or RHS of Eq.~\ref{eq:fluxMatch} can be used to set the flux at the top of the biofilm.
Note that the previous expression is well-posed if the tank boundary layer is set to zero, \ie, $L_L=0$ and the expression reduces to $S_\mathrm{top}=S_{t:k}$.

In summary, the model consists of solving Eq.~\ref{eq:singleODE}, which contains $N_\mathrm{var}$ (Eq.~\ref{eq:Nvar}) ODEs.  The RHSs of these ODEs are defined with Eqs.~\ref{eq:dXtdt}, \ref{eq:dStdt}, \ref{eq:dPbdt_discrete}, \ref{eq:dSbdt_discrete}, and \ref{eq:dLfdt}.   The next section describes the code organization that solves these equations including details on the inputs and outputs of the model. 

\section{Organization and Execution of the Code}\label{sec:org_exc}

The model is written entirely Julia.  It was developed in Julia v1.8 and tested on v1.7.  To use the model, a user will, typically, create a case file that defines the input parameters, calls the solver, and does any post-processing such as making plots or animations.  Each of these is described in more detail below.  Unit tests that compare simulated results to analytic solutions and a series of simulations compared to precomputed solutions are included and described in \ref{sec:unitTests}.  These unit tests ensure the accuracy of the code base.   

\subsection{Installation}
Installation instructions are provided in the README file with additional detailed instructions provided at \\ https://markowkes.github.io/Biofilm.jl/installation.

\subsection{Input Parameters}
\label{sec:input}
The input parameters describe the biofilm to be simulated and provide code options.  All the parameters are added to a dictionary (key:value structure).  The dictionary is initially created with
\begin{lstlisting}
# Create empty dictionary to hold parameters 
d = createDict()
\end{lstlisting}
   Parameters are then added by calling \lstinline{addParam!}.  For example, adding the simulation title and time to the dictionary can be done with, \eg,
  \begin{lstlisting}
addParam!(d, "Title",    "Single Substrate and Particulate Case")
addParam!(d, "tFinal",   1.0) 
  \end{lstlisting}
  This process of calling \lstinline{addParam!} is repeated until all the parameters are added to the dictionary.  
  The dictionary is now checked to make sure all the parameters are provided and packaged into a structure \lstinline{p} that the solver is expecting by calling 
  \begin{lstlisting}
# Package and check parameters 
p = packageCheckParam(d)
  \end{lstlisting}
  The structure \lstinline{p} fully describes the case and is the only variable passed to the solver.  It might seem redundant to have the parameters entered into the dictionary \lstinline{d} and then packaged into the struct \lstinline{p}.  While you can directly create the struct \lstinline{p}, creating the dictionary first and then the struct is recommended because when building the dictionary much more helpful error messages can be produced to help the user identify and fix issues with the parameters, then packaging the parameters into the struct allows the solver to run more efficiently. 

  Below is a complete list of each parameter.  Additional examples of setting the parameters can be found in the examples (\url{https://markowkes.github.io/Biofilm.jl/examples/}).  
  
\subsubsection*{Simulation Parameters}
\begin{itemize}
    \setlength\itemsep{0em}
    \item \lstinline{Title} : Description of the case, used, e.g., on the title of plots.
    \item \lstinline{tFinal} : Simulation is performed from $t=0$ to $t=$\lstinline{tFinal} days.
    \item \lstinline{tol} : Tolerance used for differential equation solver.  The solution will have an error less than \lstinline{tol}.
    \item \lstinline{outPeriod} : Period in days between text outputs during simulation.  Note that the solver will take smaller timesteps than \lstinline{outPeriod} to achieve the specified tolerance \lstinline{tol} and the entire solution will be available when the solver completes.
\end{itemize}

\subsubsection*{Particulate Parameters}
\begin{itemize}
    \setlength\itemsep{0em}
    \item \lstinline{XNames} : Vector of strings with the name of each particulate used on text output and plots.
    \item \lstinline{Xto} : Vector of initial particulate concentration(s) in the tank in g/m$^3$.
    \item \lstinline{Pbo} : Vector of initial particulate volume fractions(s) in the biofilm.  This value is used to set the volume fraction at all the grid cells.
    \item \lstinline{rho} : Vector of particulate densities in g/m$^3$.
    \item \lstinline{Kdet} : Detachment coefficient of particulates from the top of the biofilm in 1/m-d.
    \item \lstinline{mu} : Vector of functions that provide the particulate growth rates in 1/d. For example, in the example in Section~\ref{sec:Case3} there are two particulates, ``Live'' and ``Dead'', and one substrate. The growth rates are defined as 
    \begin{lstlisting}
addParam!(d, "mu", [(S,X,Lf,t,z,p) -> (mumax * S[1]) / (KM + S[1]) 
                    (S,X,Lf,t,z,p) -> 0.0 ] )
\end{lstlisting}
    which sets the growth rate of the first particulate to the Monod equation~\cite{monod_growth_1949}
    \begin{equation}
        \label{eq:growth rateExample}
        \mu_1=\mu_\mathrm{Live} = \mu_\mathrm{max} \frac{S_{1}}{K_M + S_{1}}
    \end{equation}
    and the second growth rate to $\mu_2=\mu_\mathrm{Dead} = 0$.  The growth rates can be other functions such as a double Monod equation, inhibition model, or any other function of the substrate concentrations \lstinline{S}, particulate concentrations \lstinline{X}, biofilm thickness \lstinline{Lf}, time \lstinline{t}, height in biofilm \lstinline{z}, or other parameters within the struct \lstinline{p}. 
    \item \lstinline{srcX} : Vector of functions that provide the sources of particulate concentrations in g/m$^3$-d.
    
\end{itemize}

\subsubsection*{Substrate Parameters}
\begin{itemize}
    \setlength\itemsep{0em}
    \item \lstinline{SNames} : Vector of strings with the name of each substrate used on text output and plots.
    \item \lstinline{Sto} : Vector of initial substrate concentration(s) in the tank in g/m$^3$.
    \item \lstinline{Sbo} : Vector of initial substrate concentration(s) in the biofilm in g/m$^3$.  This concentration is used to set the concentration at all the grid cells.
    \item \lstinline{Sin} : Vector of functions that provide the influent substrate concentrations in g/m$^3$.
    \item \lstinline{srcS} : Vector of functions that provide the sources of substrates concentrations in g/m$^3$-d.
    \item \lstinline{Yxs} : Matrix of biomass yield coefficients on the substrate.  The matrix has size $N_x \times N_s$. Within the matrix, the value in the $j^\mathrm{th}$ row and $k^\mathrm{th}$ column, \ie, \lstinline{Yxs[j,k]}, provides the number of grams of the $j^\mathrm{th}$ particulate grown for each gram of the $k^\mathrm{th}$ substrate consumed.  Positive values of \lstinline{Yxs} indicate substrate consumption and negative values indicate substrate generation by particulate growth.  If the particulate does not depend on the substrate the yield coefficient should be infinite.  However, to make this input more user-friendly, a value of \lstinline{0} or \lstinline{Inf} can be entered and the code automatically converts a \lstinline{0} to \lstinline{Inf}. 
    \item \lstinline{Db} : Vector of effective substrate diffusion coefficients through the biofilm in m$^2$/day. 
    \item \lstinline{Dt} : Vector of aqueous substrate diffusion coefficients in the fluid environment within the tank in m$^2$/day.
\end{itemize}

\subsubsection*{Tank Parameters}
\begin{itemize}
    \setlength\itemsep{0em}
    \item \lstinline{V} : Volume of tank in m$^3$.
    \item \lstinline{A} : Area of biofilm within the tank in m$^2$.
    \item \lstinline{Q} : Flowrate into and out of tank in m$^3$/day.
\end{itemize}

\subsubsection*{Biofilm Parameters}
\begin{itemize}
    \setlength\itemsep{0em}
    \item \lstinline{Nz} : Number of grid points used to discretize the biofilm.
    \item \lstinline{LL} : Thickness of boundary layer within the tank on the surface of biofilm in m.
    \item \lstinline{Lfo} : Initial thickness of the biofilm in m.
\end{itemize}

\subsubsection*{Optional Parameters}
\begin{itemize}
    \setlength\itemsep{0em}
    \item \lstinline{plotPeriod} [default=\lstinline{outPeriod}] : Period in days between plot renderings during the simulation.  For long simulations, you can increase \lstinline{plotPeriod} to reduce the number of plots that are created and speed up the simulation. Note: \lstinline{plotPeriod} is required to be a multiple of \lstinline{outPeriod}.
    \item \lstinline{plotSize} [default=\lstinline{(1600,1000)}] : Size of plots in pixels.
    \item \lstinline{makePlots} [default=\lstinline{true}] : Boolean that controls if code should produce plots while running.  
    \item \lstinline{optionalPlot} [default=\lstinline{"growthrate"}] : While running, the code displays six plots.  The sixth plot defaults to plotting the particulate growth rate versus location in biofilm.  Setting \lstinline{optionalPlot = "source"} will change this plot to the particulate source term versus location in biofilm.
\end{itemize}

When defining the parameters it is often useful to reference other variables or functions.  For example, when defining the growth rate in Eq.~\ref{eq:growth rateExample} the variables $\mu_\mathrm{max}$ and $K_m$ are used.  In the code, these variables can be defined before \lstinline{growthrate} is added to the dictionary. Similarly, functions can be defined and used when defining parameters, see Example~\ref{example:phototroph}.

\subsection{Overview of Solver}
The solver consists of a collection of Julia (.jl) files.  Each file is described in additional detail below.

~\\[-0.1em]\textit{Biofilm.jl} :\quad
is the main file of the code base and defines the Biofilm module, exported functions and objects, and includes all the other files. Biofilm.jl is used by first including the module
\begin{lstlisting}
julia> using Biofilm 
\end{lstlisting}
and then calling one of the functions described below.

~\\[-0.1em]\textit{solver.jl} :\quad
contains the function \lstinline{BiofilmSolver(p::param)}, which runs a simulation of a biofilm using the parameters provided in the struct \lstinline{p} that includes all the inputs described in Section~\ref{sec:input}.    This function prepares the solver, runs the solver, and formats and returns the solution to be easily interpreted by the user.

The workhorse within \lstinline{BiofilmSolver} is a call to the function \lstinline{solve()}, which is part of the DifferentialEquations.jl package~\cite{rackauckas_differentialequationsjl_2017}  This package contains many state-of-the-art methods for solving differential equations and is key to the speed of Biofilm.jl.  The ODE solver computes the solution of Eq.~\ref{eq:singleODE}.

~\\[-0.1em]\textit{rhs.jl} :\quad
contains \lstinline{biofilmRHS!()}, which computes the RHS of the differential equations in Eq.~\ref{eq:singleODE}.  The first part of this function splits the vector of dependent variables $\bm{\mathrm{sol}}$ into $X_t$, $S_t$, $\bm{P}_b$, $\bm{S}_b$, and $L_f$.  Next, a number of intermediate variables are computed which are terms that appear multiple times in the RHSs by calling functions that are defined in computes.jl.  Finally the RHS of Eqs.~\ref{eq:dXtdt}, \ref{eq:dStdt}, \ref{eq:dPbdt_discrete}, \ref{eq:dSbdt_discrete}, and \ref{eq:dLfdt} are computed and organized in a single vector.

~\\[-0.1em]\textit{computes.jl} :\quad
contains functions that compute intermediate variables that appear multiple times in the evaluation of the differential equation RHSs.  By computing these variables once the results can be reused to improve the speed of the code. 

~\\[-0.1em]\textit{outputs.jl} :\quad
produces a text output to the terminal and plots the solution.

~\\[-0.1em]\textit{structs.jl} :\quad
defines the composite types or structs used by the code.  These structs hold a number of parameters or variables together to reduce the number of variables that need to be passed into functions.  At the top of each function, any variables from the structures are accessed with the \lstinline{@unpack} macro.

~\\[-0.1em]\textit{parameters.jl} :\quad
defines the functions \lstinline{createDict()}, \lstinline{addParam!(d,name,value)}, and \lstinline{packageCheckParam(d)} that are used to create a dictionary of parameters and package and check the parameters.  Additionally, this file provides the function \lstinline{printDict(d)} that can be used to print all the values in the a dictionary.

~\\[-0.1em]\textit{tools.jl} :\quad
provides a variety of helper functions.

~\\[-0.1em]\textit{postprocess.jl} :\quad
contains functions to help a user analyze the results of a simulation.  Additional details are provided in the next section. 

\subsection{Postprocessing Output}
A simulation is run by calling
\begin{lstlisting}
julia> t,zm,Xt,St,Pb,Sb,Lf,sol = BiofilmSolver(p) # Run solver
\end{lstlisting}
The results are returned in the following output variables.
\lstinline{t} is a vector of the solution times.
\lstinline{zm} is a vector of grid cell centers used to discretize the biofilm. Since the biofilm thickness (and thus the grid) changes with time, only the grid at the final time is provided as an output.  
\lstinline{Xt} and \lstinline{St} are arrays of the particulate and substrate concentrations in the tank as a function of times \lstinline{t}.
The variables \lstinline{Pb} and \lstinline{Sb} are arrays of the biofilm particulate volume fractions and substrate concentrations at the final simulation time as a function of the grid point locations \lstinline{zm}.
Finally, \lstinline{sol} is an object that is created by DifferentialEquations.jl that contains the complete solution of the ODE solver.  Although it contains the entire solution, it is difficult to process since it contains $N_\mathrm{var}$  (Eq.~\ref{eq:Nvar}) dependent variables. 
These outputs can be analyzed to provide additional information on the biofilm that has been simulated using the following functions.

~\\[-0.1em]\lstinline{biofilm_plot(sol,p)} :\quad can be used to produce the standard plots of the simulation as shown in, \eg, Fig.~\ref{fig:Case1}.  This is useful to make a plot of the solution at the end of the simulation, especially if plots are turned off during the simulation using the optional parameter \lstinline{makePlots=false}.  

This function calls a plotting recipe and allows for considerable customization of the plot using the inputs of the Plots.jl library and solution processing form DifferentialEquations.jl.  The examples in Section~\ref{sec:examples} demonstrate some plots that can be created with this function. 

~\\[-0.1em]\lstinline{biofilm_analyze(sol,p,times [,makePlot=true])} :\quad will display the dependent variables at the specified times \lstinline{t}.  For example, after running the example described in Section~\ref{sec:Case1}, the solution could be analyzed to check the solution at $t=[0,0.25,0.5,0.75,1.0]$, \ie,
\begin{lstlisting}
julia> times = 0:0.25:1.0
julia> biofilm_analyze( sol, p, times)
Analyzing Single Substrate and Particulate Case
   Time   |      Bug |   Oxygen | min,max(     Bug) | min,max(  Oxygen) |       Lf 
    0.000 |       10 |       10 |     0.08,    0.08 |        0,       0 |       10
    0.250 |      102 |     51.7 |     0.08,    0.08 |     29.1,    51.3 |      545
    0.500 |      256 |     2.94 |     0.08,    0.08 |    0.568,    2.87 |      348
    0.750 |      257 |     2.93 |     0.08,    0.08 |    0.745,    2.87 |      312
    1.000 |      257 |     2.93 |     0.08,    0.08 |    0.761,    2.87 |      309
\end{lstlisting}
where the first column is time, the second and third columns are the particulate and substrate concentrations in the tank, the fourth and fifth columns are the minimum and maximum of the particulate volume fractions and substrate concentration within the biofilm, and the last column is the biofilm thickness.  This function provides the outputs \lstinline{Xt}, \lstinline{St}, and \lstinline{Lf} at the specified times for further postprocessing. For example, this is used to process the example in Section~\ref{sec:Case5} to create Fig.~\ref{fig:Case5_2days}. 

Including the optional parameter \lstinline{makePlots=true} will create a plot of the solution within the biofilm at each specified time.  This is used in the processing of the example in Section~\ref{sec:Case5}.  

~\\[-0.1em]\lstinline{biofilm_movie(sol, p, times [, filename="anim.gif", fps=20)]} :\quad can be used to make an animation of the solution at the specified times.  The optional parameters \lstinline{filename} and \lstinline{fps} can be included to change the filename or frame rate of the movie.  This function uses the \lstinline{@animate} macro and \lstinline{gif()} function from the Plots.jl package. 

~\\[-0.1em]\lstinline{biofilm_sol2csv( sol, p [, filename="biofilm.csv"])} :\quad takes the output of a simulation and writes the results as a CSV file.  Each row in the file corresponds to a different output time and the column contains the time, the tank particulate and substrate concentrations, biofilm particulate volume fractions and substrate concentrations, and finally the biofilm thickness.  This function can be useful for users that prefer looking at results in a spreadsheet.

\section{Examples}\label{sec:examples}
This section provides a series of example cases that highlight the capabilities of Biofilm.jl.  These examples are the same as those provided with the code, however, additional postprocessing has been done for some of the cases.  The code used to create the results in this manuscript is available in a branch called CodePaper accessible at \url{https://github.com/markowkes/Biofilm.jl/tree/CodePaper/examples}

\subsection{Case 1: Single Substrate and Particulate}\label{sec:Case1}
\begin{figure}[htbp]
    \centering
    \includegraphics[width=0.9\textwidth]{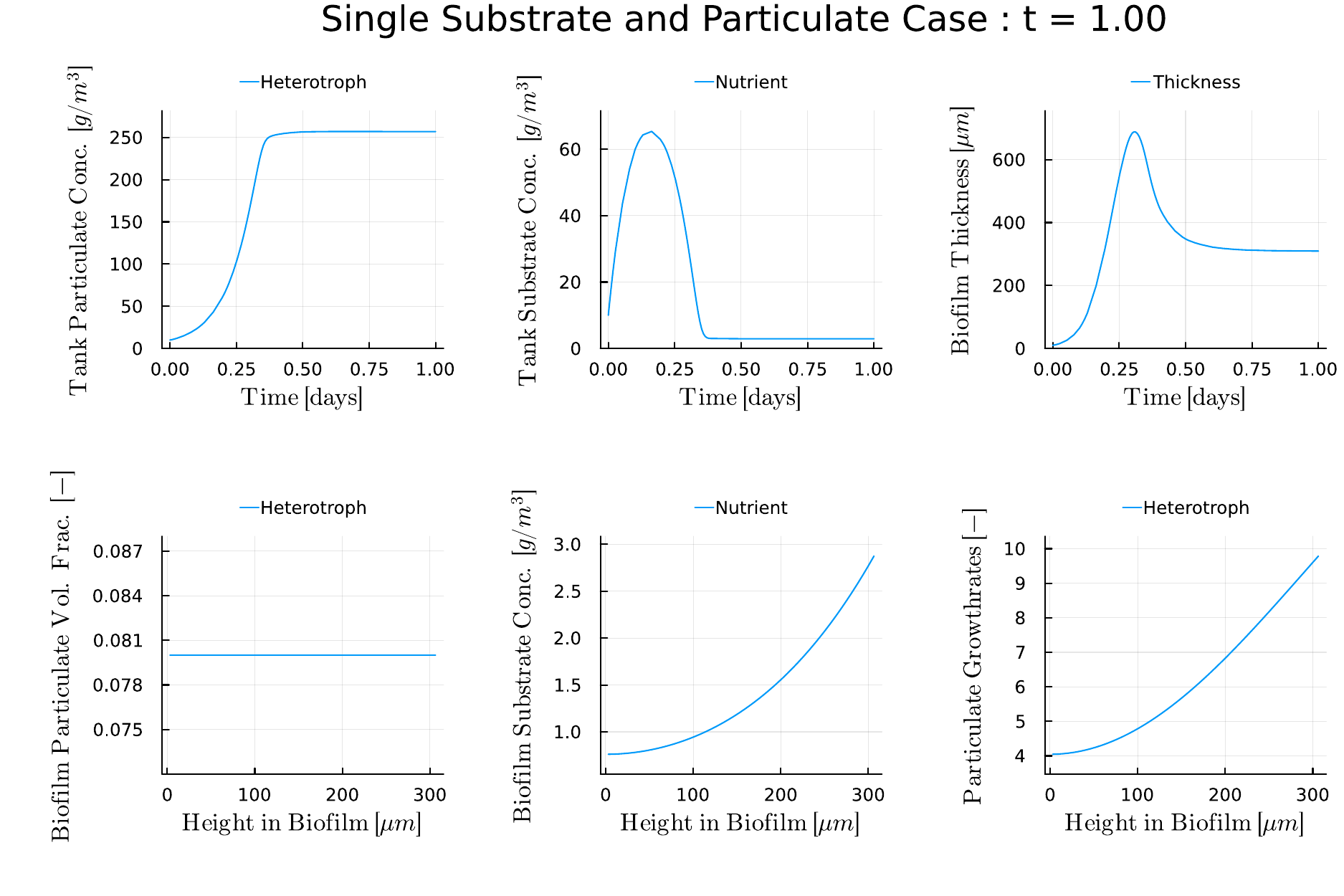}
    \caption{Standard output at the end of running Case1.jl.}
    \label{fig:Case1}
\end{figure}

This is a simple example of a single particulate named \lstinline{"Heterotroph"} that consumes the substrate \lstinline{"Nutrient"} to grow using the Monod equation \cite{monod_growth_1949}.
The case is run by starting Julia, installing Biofilm.jl, and executing
\begin{lstlisting}
julia> include("examples/Case1.jl")
\end{lstlisting}
The output from the program is shown in Fig.~\ref{fig:Case1}.  Note that the size of the plot was adjusted by creating the plot with the command
\begin{lstlisting}
julia> biofilm_plot(sol,p,size=(900,600))
\end{lstlisting} 
which is run after the simulation is complete. The top row shows how the particulate concentration (left), substrate concentration (center), and biofilm thickness (right) vary with time.  The bottom row shows the particulate volume fraction (left), substrate concentration (center), and particulate growth rate (right) versus location within the biofilm at the end of the simulation (1.00 days for this case).

The heterotroph concentration in the tank starts at the initial condition of \lstinline{Xto}=10 g/m$^3$ and increases to a steady-state value of 257 g/m$^3$ as seen in the top-left plot.  The nutrient concentration in the tank starts at an initial concentration of \lstinline{Sto}=10 g/m$^3$, increases quickly due to the flow into the tank with a concentration of \lstinline{Sin}=100 g/m$^3$, then starts decreasing as the heterotroph concentration increases and more of the nutrient is consumed until a steady-state of 2.92 g/m$^3$ is reached as shown in the top-center plot.  The biofilm thickness increases from a value of \lstinline{Lfo}=1.0E-5 $\mu$m quickly due to growth in the initial system with an abundance of the nutrient. Eventually, the thickness decreases to a steady thickness of 309 $\mu$m as observed in the top-right plot.

Within the biofilm at the final time of \lstinline{t}=1.0 days, the bottom-left plot shows that the heterotroph volume fraction is 0.08 everywhere.  This value comes from the initial condition of \lstinline{Pbo}=0.08 and the fact that there is only one particulate in this simple system. The nutrient concentration, in the bottom-center plot, is highest at the top of the biofilm and decreases within the biofilm as diffusion of the nutrient into the biofilm is balanced by the consumption by the heterotroph.  The concentration of the nutrient at the top of the biofilm has a value of \lstinline{Sb[Nz]}=2.87 g/m$^3$, which is close to the steady-state tank concentration of 2.92 g/m$^3$.  The difference is due to the diffusion rate through the boundary layer and very top of biofilm (Eq.~\ref{eq:fluxMatch})   The heterotroph's growth rate, shown in the bottom-right plot, is highest near the top of the biofilm where the nutrient concentration is highest and decreases further into the biofilm.

\subsection{Case 2: Multiple Substrates}\label{sec:Case2}
This example is based on the acid stress response in Zhang et al.~\cite{zhang_general_2013} and features biomass $X$ that consumes glucose $s$ and produces lactate $p$.  Here the variables $s$ and $p$ stand for substrate and product, respectively.  Note that both $s$ and $p$ are modeled with Eqns.~\ref{eq:dStdt} and~\ref{eq:dSbdt} with $S=[s,p]^\mathsf{T}$. 

The presence of lactate can inhibit the growth of biomass.  This appears as a reduction in growth rate, as $p$ increases, until a maximum value of $p_\mathrm{max}$ above which the growth ceases, \ie,
\begin{equation}
    \mu = \begin{cases}
        \mu_\mathrm{max} s \left(1 - \displaystyle\frac{p}{p_\mathrm{max}}\right) & p \le p_\mathrm{max}\\
        0.0 & p> p_\mathrm{max}
        \end{cases}
\end{equation}
where the term in parentheses captures inhibition.  In the code, this is done with two steps.  The first is to define a function that computes the growth rate for a single combination of $s$ and $p$, \ie,
\begin{lstlisting}
# Define mu function 
function mu(s,p)
    if p < p_max 
        mu = mu_max*s*(1-p/p_max)
    else
        mu = 0.0
    end
    return mu
end
\end{lstlisting}
Next, this function is provided to the growth rate variable in the parameter dictionary (see Section~\ref{sec:input}), \ie, 
\begin{lstlisting}
# Growthrate: call mu(s,p) for S = [s,p]
addParam!(d, "mu", [(S,X,Lf,t,z,p) -> mu(S[1],S[2])])
\end{lstlisting}
which calls \lstinline{mu(s,p)} function using the notation \lstinline{S = [S[1],S[2]] = [s,p]}.

\begin{figure}[htbp]
    \centering
    \includegraphics[width=0.9\textwidth]{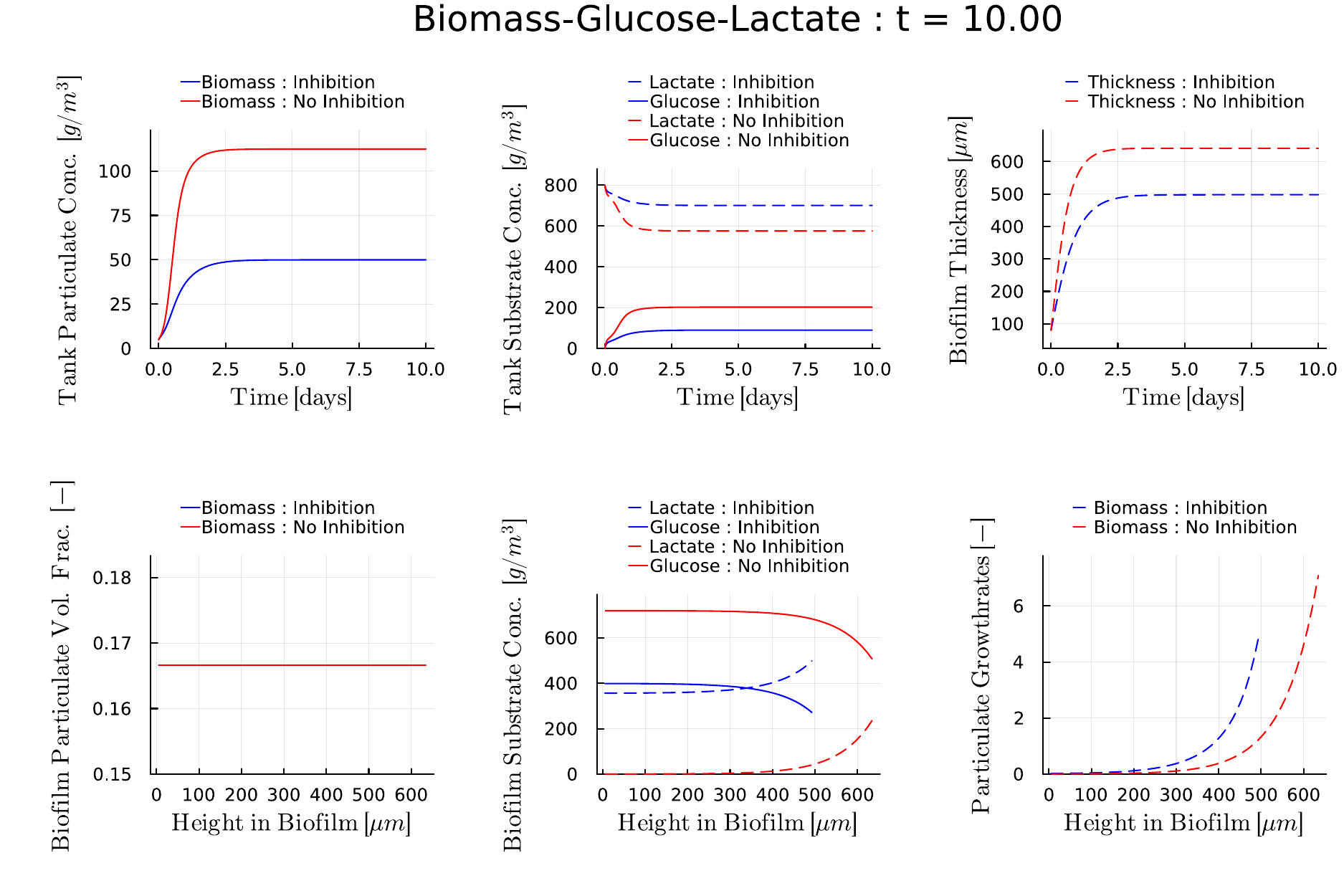}
    \caption{Output of Case2.jl run with (blue) and without (red) growth inhibition.}
    \label{fig:Case2_noinhibition}
\end{figure}

To see the impact of the inhibition, two simulations are run with the inhibition turned on and off, respectively. The results from both simulations are shown in Fig.~\ref{fig:Case2_noinhibition} which was generated with the code 
\begin{lstlisting}
# Run with inhibition 
p_max = 400
t_in,zm_in,Xt_in,St_in,Pb_in,Sb_in,Lf_in,sol_in = BiofilmSolver(p)
plt = biofilm_plot(sol_in,p,"Inhibition",
                   size=(900,600), line=(:blue,[:solid :dash]))

# Run without inhibition & add to plot
p_max  = Inf  
t_no,zm_no,Xt_no,St_no,Pb_no,Sb_no,Lf_no,sol_no = BiofilmSolver(p)
plt = biofilm_plot!(plt,sol_no,p,"No Inhibition",
                    size=(900,600), line=(:red,[:solid :dash]))
\end{lstlisting}
This code is run after running a standard Case 3 simulation which sets the case parameters.  The code above runs two simulations, the first with the inhibition turned on (\lstinline{p_max = 400}) and plots the results in blue.  Next a second simulation is run without inhibition (\lstinline{p_max = Inf} and the result is added to the previous plot and colored in red. Note that the second plot command uses \lstinline{biofilm_plot!} (note the exclamation point), which adds to the plot instead of creating a new plot.  

The results in Fig.~\ref{fig:Case2_noinhibition} show the output of both simulations.  In the top row, the biomass (left) and substrate (center) concentrations along with the biofilm thickness (right) are shown versus time. After roughly 2.5 days all the quantities reach a steady state.  The bottom-left plot shows the biomass volume fraction, which since there is only one particulate remains at the input value of 1/6 throughout the biofilm.  The bottom-center plot shows the glucose and lactate concentrations throughout the biofilm.  The glucose (solid lines) is diffusing from the tank into the biofilm and is consumed by the biomass, which lowers the concentration at lower heights.  The lactate (dashed lines) is produced by biomass and has high concentrations in the lower part of the biofilm.  Diffusion moves the lactate toward the surface of the biofilm where it enters the tank.  The biomass growth rate is shown in the bottom-right figure.  Growth is highest at the top of the biofilm where high glucose concentrations exists and decreases at lower heights in the biofilm.

Comparing the inhibited (blue) results to the case without inhibition (red) shows that without inhibition the biomass concentrations in the tank (top-left) is over 2 times larger and the biofilm is roughly 25\% thicker (top-right).  For both cases the growth of biomass occurs at the top of the biofilm and ceases at lower heights within the biofilm, however the limit to growth is different in the two cases.  Without inhibition, the biomass consumes all of the glucose that is diffusing through the biofilm and the glucose concentration goes to zero, limiting growth, at lower heights in the biofilm (bottom-center).  With inhibition, growth is limited by the inhibition due to the higher concentrations of lactate lower in the biofilm (bottom-center) and some glucose exists even at the bottom of the biofilm (bottom-center).

\subsection{Case 3: Multiple Particulates}\label{sec:Case3}
This example demonstrates a simulation with multiple particulates, name\-ly \lstinline{"Living Bug"} and \lstinline{"Dead Bug"}.  The living bug consumes the substrate and grows using the Monod equation.  The living bugs die and turn into dead bugs, this is captured in the model with the source terms
\begin{align*}
    \mathrm{src}_\mathrm{Living} &= - b X_\mathrm{Living}\text{,~ and} \\
    \mathrm{src}_\mathrm{Dead} &= \hphantom{-} b X_\mathrm{Living}.
\end{align*}

\begin{figure}[htbp]
    \centering
    \includegraphics[width=0.9\textwidth]{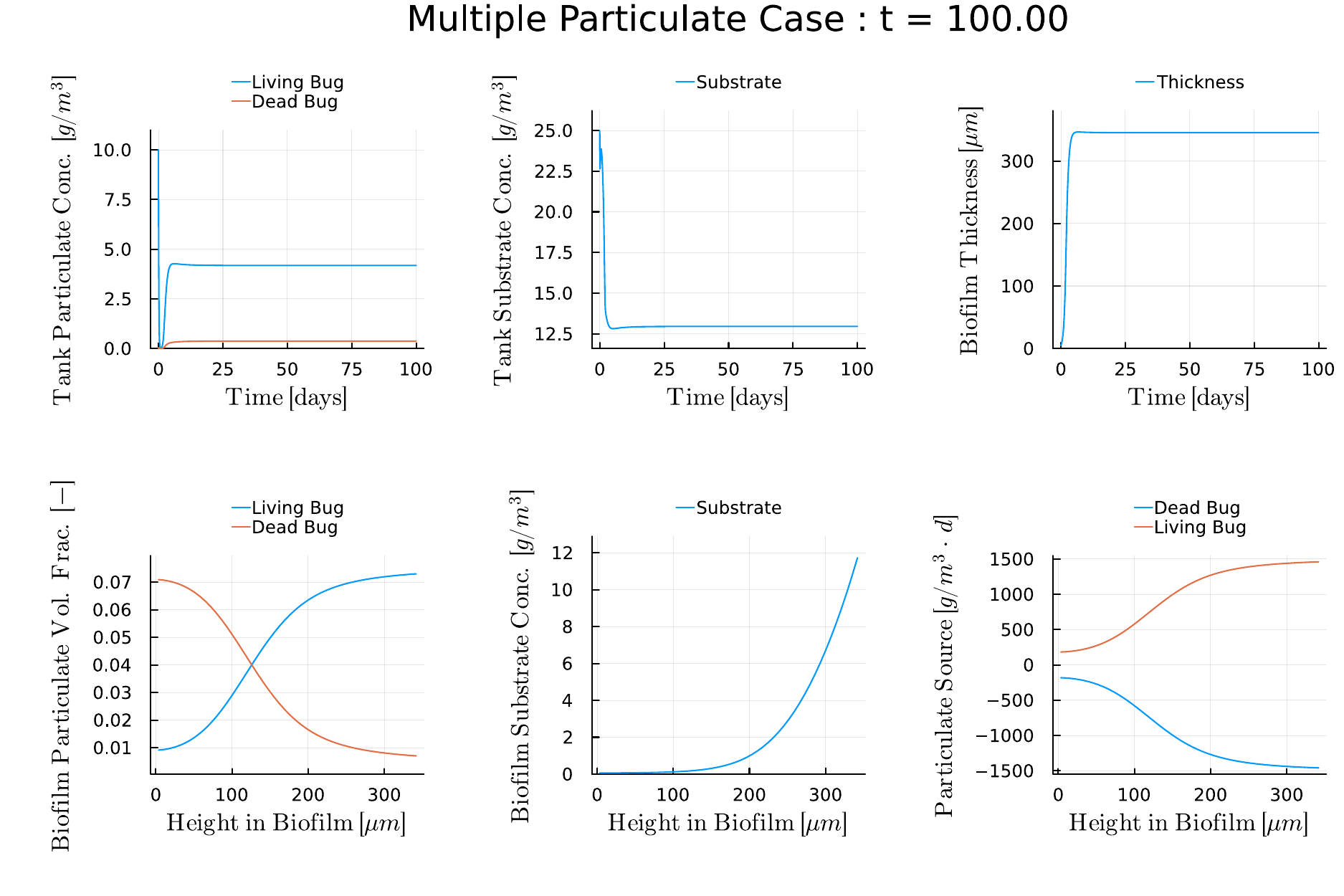}
    \caption{Standard output at the end of running Case3.jl.}
    \label{fig:Case3}
\end{figure}

The results at the end of the 100-day simulation are shown in Fig.~\ref{fig:Case3}.  Of note for this case are the variations within the biofilm.  Near the surface of the biofilm, the substrate diffuses from the tank (bottom-center plot) and is consumed by the living bugs causing the living bug volume fraction to be larger in roughly the top half of the biofilm.  Deeper in the biofilm, the living bugs do not receive substrate and die causing the dead bug volume fraction to increase.  For this case, the parameter \lstinline{optionalPlot="source"} is used by adding it to the parameter dictionary with 
\begin{lstlisting}
addParam!(d, "optionalPlot", "source")
\end{lstlisting}
With this parameter, the bottom-right figure shows the source term that transfers living bugs to dead bugs. The source has the largest magnitude near the top of the biofilm where the living bug concentration is the highest. 

\subsection{Case 4: Multiple Substrates and Particulates}\label{sec:Case4}
This case demonstrates how to model a biofilm with three substrates and three particulates.  The substrates are oxygen, sulfate, and hydrogen sulfide and the particulates are sulfide-oxidizing bacteria (SOB), sulfate-reducing bacteria (SRB), and dead bacteria.  

The system requires interactions between the bacteria and substrates to function.  
Oxygen and sulfate enter the tank with the influent. 
The SRB uses the sulfate and produces sulfide, but this growth is inhibited by oxygen.  
The SOB can then use the sulfide, produced by the SRB, and oxygen to grow.
The SRB and SOB slowly die creating the dead bacteria. 
These dynamics appear in the growth rates which can be written as
\begin{align*}
    \mu_\mathrm{SOB} &= \mu_\mathrm{max:B} \left( \frac{S_1}{K_{m:B_1} + S_1} \cdot \frac{S_3}{K_{m:B_3} + S_3} \right) \\
    \mu_\mathrm{SRB} &= \mu_\mathrm{max:C} \left( \frac{S_2}{K_{m:C_2} + S_2} \cdot \frac{1.0}{1.0 + S_1/K_I} \right) \\
    \mu_\mathrm{Dead} &= 0.0
\end{align*}
where $S_1$, $S_2$, and $S_3$ correspond to oxygen, sulfate, and hydrogen sulfide, respectively.  
The parameters $\mu_\mathrm{max:B}$, $\mu_\mathrm{max:C}$, $K_{m:B_1}$, $K_{m:B_3}$, $K_{m:C_2}$, and $K_I$ are constants and set in the case file. 
The source terms can be written as
\begin{align*}
    \mathrm{src}_\mathrm{SOB} & = - D_\mathrm{SOB} X_1\\
    \mathrm{src}_\mathrm{SRB} & = \hphantom{- D_\mathrm{SOB} X_1} - D_\mathrm{SRB} X_2 - D_\mathrm{O:SRB} S_1 \\
    \mathrm{src}_\mathrm{Dead} & = + D_\mathrm{SOB} X_1 + D_\mathrm{SRB} X_2 + D_\mathrm{O:SRB} S_1 
\end{align*}
where  $X_1$, $X_2$, and $X_3$ correspond to SOB, SRB, and dead biomass, respectively.  The constants $D_\mathrm{SOB}$ and $D_\mathrm{SRB}$ describe the death rate of SOB and SRB, respectively.  $D_{O:\mathrm{SOB}}$ is the death rate of SRB due to the precense of oxygen.

These expressions are coded as
\begin{lstlisting}
addParam!(d, "mu", [
    (S,X,Lf,t,z,p) -> mumaxB*(S[1]./(KmB1.+S[1])).*(S[3]./(KmB3.+S[3])),   # SOB
    (S,X,Lf,t,z,p) -> mumaxC*(S[2]./(KmC2.+S[2])).*(1.0./(1.0.+S[1]/KI)) , # SRB
    (S,X,Lf,t,z,p) -> 0.0 ] )                                              # Dead
\end{lstlisting}
and
\begin{lstlisting}
addParam!(d, "srcX",  [
    (S,X,Lf,t,z,p) -> - D_SOB*X[1]                             ,  # SOB
    (S,X,Lf,t,z,p) ->              - D_SRB*X[2] - D_O_SRB*S[1] ,  # SRB
    (S,X,Lf,t,z,p) -> + D_SRB*X[2] + D_SOB*X[1] + D_O_SRB*S[1] ]) # Dead
\end{lstlisting}
    
\begin{figure}[htbp]
    \centering
    \includegraphics[width=0.9\textwidth]{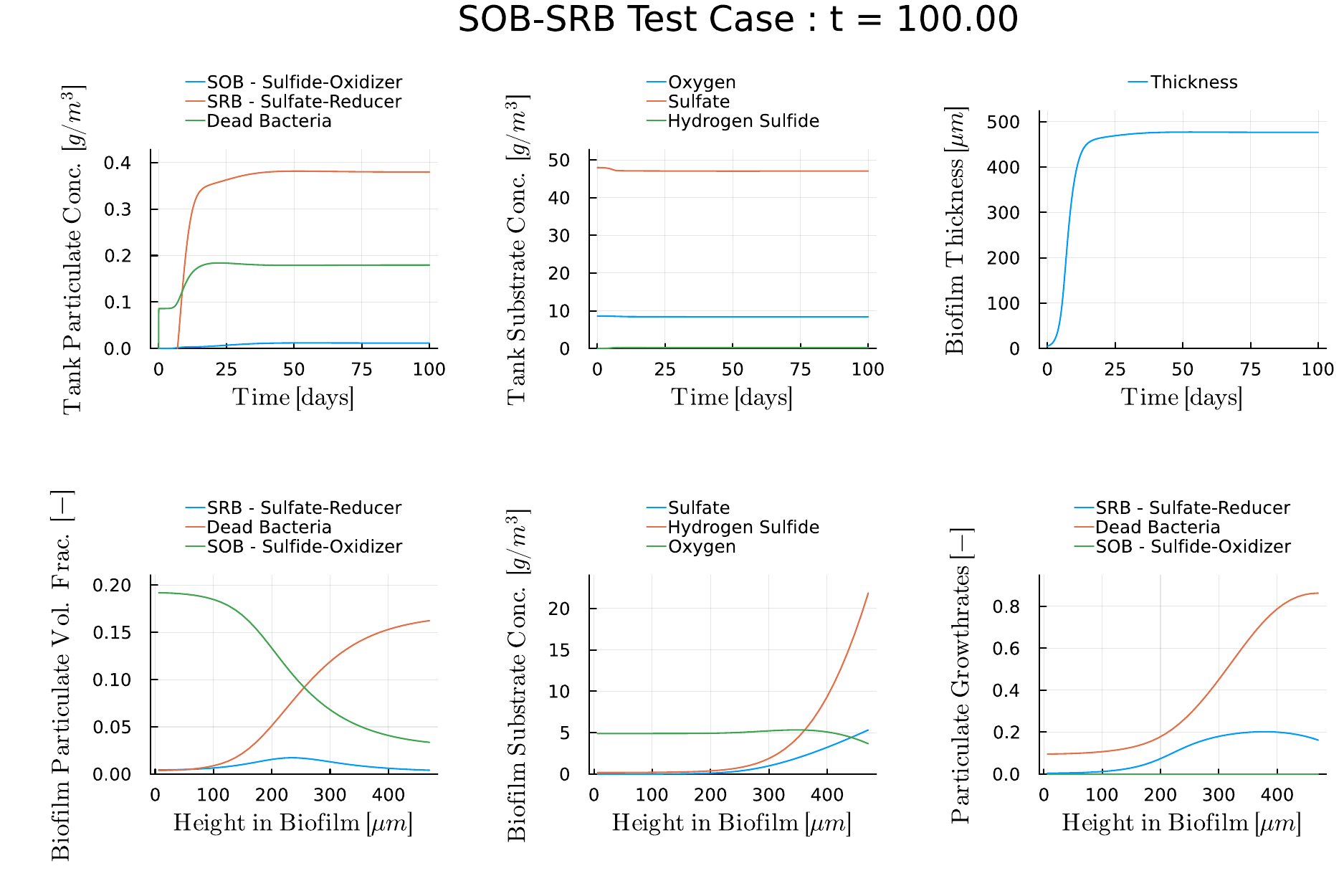}
    \caption{Standard output at the end of running Case4.jl.}
    \label{fig:Case4}
\end{figure}

The results after 100 days are shown in Fig.~\ref{fig:Case4}.  The top row shows the SOB, SRB, and dead bacteria concentrations within the tank (left), substrate concentrations (center), and biofilm thickness (right).  After roughly 50 days the quantities reach a steady state.  The bottom row shows the particulate volume fractions (left), substrate concentrations (center), and particulate growth rates (right).  In the bottom of the biofilm, a layer of dead cells collects as little sulfate (needed by SRB) or oxygen (needed by SOB) reaches this layer.  In the middle region, the SOB lives and has access to the sulfide from the SRB and oxygen diffusing from the tank.  In the top layer, the SRB thrives as the sulfate concentration is very high. 

\subsection{Case 5: Phototroph}\label{sec:Case5}
\label{example:phototroph}
This example demonstrates how a particulate that grows in light with a diurnal cycle can be simulated.  The particulate is a phototroph with a growth rate that depends on light intensity $I(t,z)$, \ie,
\begin{equation*}
    \mu_\mathrm{phototroph} = \mu_\mathrm{max} I(t,z).
\end{equation*}
As the phototroph grows it generates the substrate, oxygen, through the biomass yield coefficient \lstinline{Yxs=[-0.52]}.

\begin{figure}[htbp]
    \centering
    \begin{subfigure}[b]{0.4\textwidth}
        \centering
        \includegraphics[width=\textwidth]{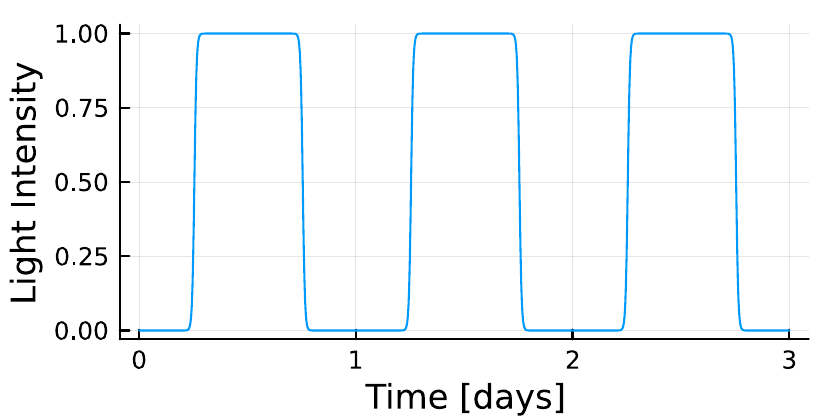}
        \caption{Variation with time}
        \label{fig:lightTime}
    \end{subfigure}
    \quad
    \begin{subfigure}[b]{0.4\textwidth}
        \centering
        \includegraphics[width=\textwidth]{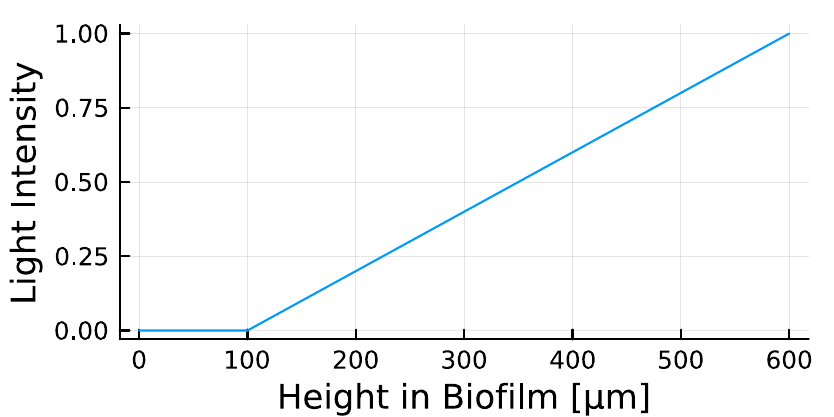}
        \caption{Variation with the location in biofilm}
        \label{fig:lightDepth}
    \end{subfigure}
    \caption{Light intensity versus time and location within biofilm for Case 5.}
    \label{fig:light}
\end{figure}

The light within the tank turns on and off throughout each day.  Mathematically, the intensity of light within the tank is 1.0 on each day after 0.25 days and then 0.0 at 0.75 days creating the on/off cycle.  In the code, this is achieved by defining the intensity function as
\begin{lstlisting}
# Define smoothed Heaviside function
smoothHeaviside(t,t0)=0.5*tanh.(100*(t.-t0).-0.5).+0.5
# Define light intensity as function of time
intensity(t) = 1.0 - (smoothHeaviside(mod(t,1),0.25)
                     -smoothHeaviside(mod(t,1),0.75))
\end{lstlisting}
where the \lstinline{smoothHeaviside} function spreads out the discontinuity in intensity over a finite time.
Furthermore, the light dissipates as it travels through the biofilm and the dissipation is defined with
\begin{lstlisting}
diss=2000; # Dissipation rate [1/m]
dissipation(z,Lf) = max.(0.0,1.0.-(Lf.-z)*diss)
\end{lstlisting}
Finally, the light intensity at a specified time and position in the biofilm can be computed with
\begin{lstlisting}
# Define light intensity as a function of
# time and location within biofilm
light(t,z,Lf) = intensity(t)*dissipation(z,Lf)
\end{lstlisting}
These functions are defined before the input parameters and are called when the growth rate is defined as
\begin{lstlisting}
mumax = 0.4;
addParam!(d, "mu", [(S,X,Lf,t,z,p) -> mumax*light(t,z,Lf)])
\end{lstlisting}
To check the light intensity function, we can make a plot of light intensity over a few days within the tank, \eg,
\begin{lstlisting}
# Define constants
Lf = 600e-6; z = Lf;
# Define times to plot
t = range(0.0,3.0,1000)
# Make plot
plot(t,map(t -> light(t,z,Lf),t))
\end{lstlisting}
which produces the plot shown in Fig.~\ref{fig:lightTime}.  Figure~\ref{fig:lightDepth} shows how the light dissipates as it travels through the biofilm and it is largest at the top of the biofilm (600 $\mu$m) and decreases linearly until reaching a zero intensity.

Due to the diurnal discontinuities in the light intensity, it is important to make sure the solver reduces the timestep near the discontinuity.  This is achieved by setting \lstinline{discontinuityPeriod=0.25} using 
\begin{lstlisting}
# Let the solver know when discontinuities (changes in light) occur
addParam!(d, "discontinuityPeriod",0.25)  
\end{lstlisting}
which makes the solver look for discontinuities every 0.25 days.  Without this parameter, periods when the light is turned on or off are missed since the solver uses large timesteps. 

\begin{figure}[htbp]
    \centering
    \includegraphics[width=0.9\textwidth]{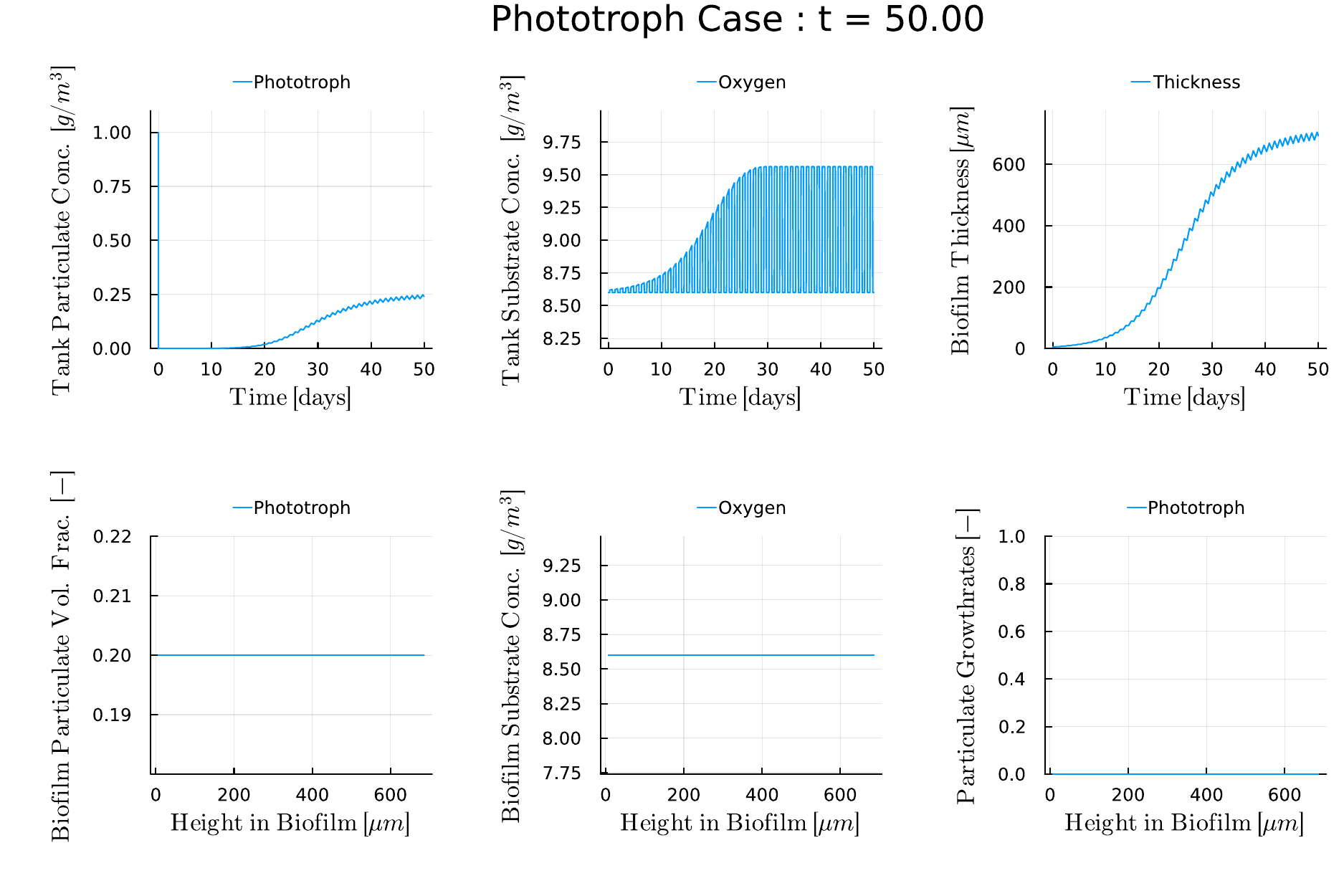}
    \caption{Standard output at the end of running Case5.jl.}
    \label{fig:Case5}
\end{figure}
The output of this example is shown in Fig.~\ref{fig:Case5}.  The top row shows the phototroph concentration (top-left), oxygen concentration (top-center), and biofilm thickness (top-right) all as a function of time.  The results vary during each day as the light turns on and off.  The bottom row displays the variations within the biofilm at the final time of 50 days when the light is off.  With the light off, no growth is occurring and the quantities do not vary throughout the biofilm.

\begin{figure}[htbp]
    \centering
    \includegraphics[width=0.9\textwidth]{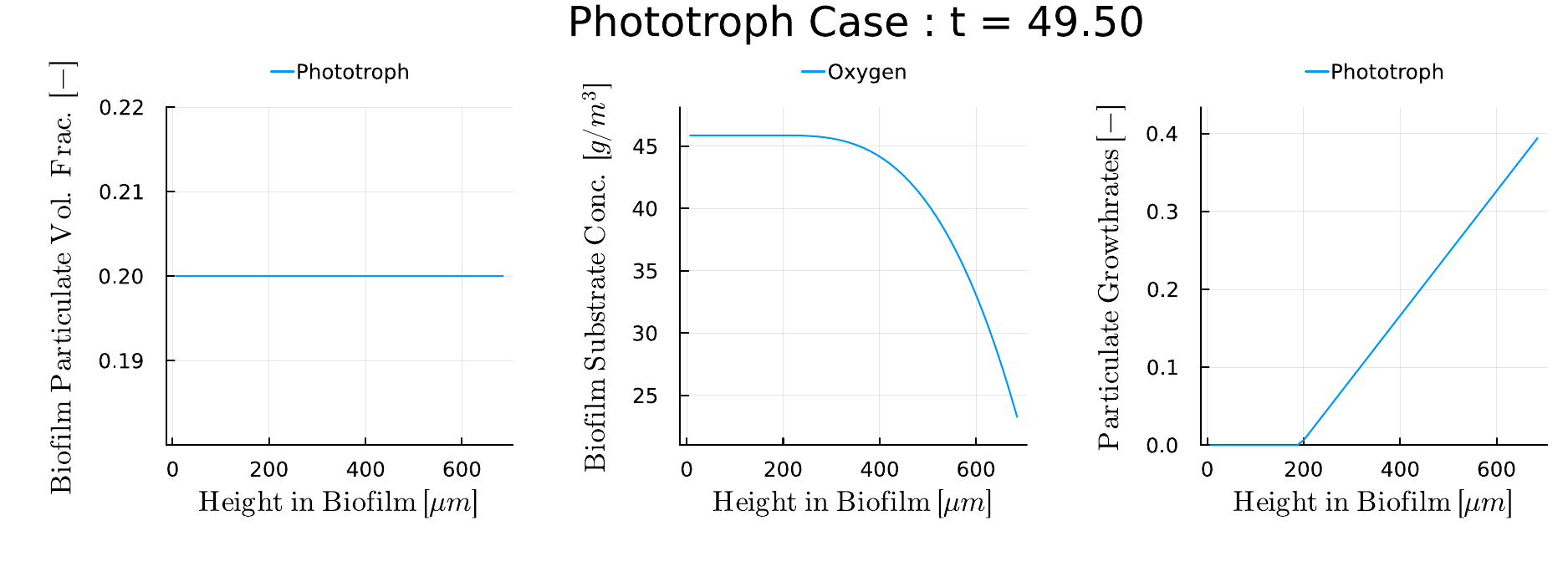}
    \caption{Analysis of Case 5 solution at t=49.5 days.}
    \label{fig:Case5_lighton}
\end{figure}
The quantities within the biofilm can be analyzed at different times. In particular, it will be useful to analyze the results when the light is on.  This can be done by running the \lstinline{biofilm_analyze()} postprocessing function with the solution of Case 5 when the light is on (\eg, 49.5 days) which is done by running 
\begin{lstlisting}
tout=49.5 # Time when the light is on
biofilm_analyze(sol,p,tout,makePlot=true)
\end{lstlisting}
This code analyzes the results and produces the plots shown in Fig.~\ref{fig:Case5_lighton}. When the light is on, the phototroph is growing (right) and the oxygen concentration is high within the biofilm as it is produced by the phototroph and decreases towards the top of the biofilm due to diffusion (center).  Since there is only one particulate, the figure of the volume fraction (left) remains the same throughout the entire simulation.

\begin{figure}[htbp]
    \centering
    \begin{subfigure}[b]{0.25\textwidth}
        \centering
        \includegraphics[width=\textwidth]{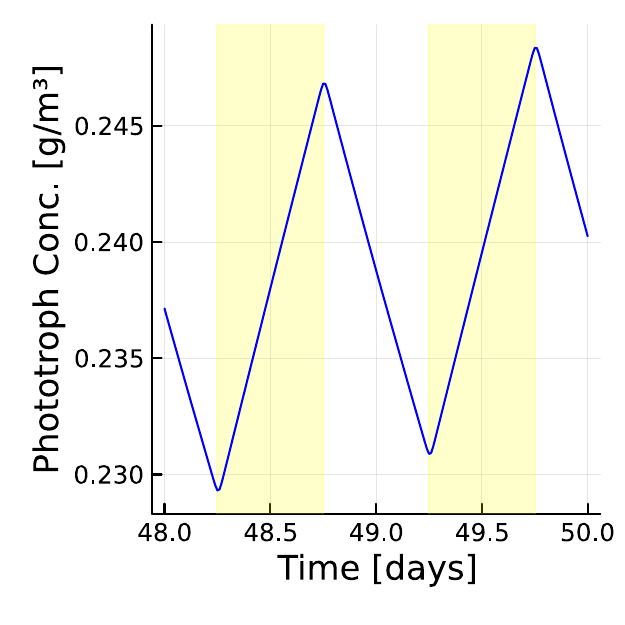}
    \end{subfigure}
    \quad
    \begin{subfigure}[b]{0.25\textwidth}
        \centering
        \includegraphics[width=\textwidth]{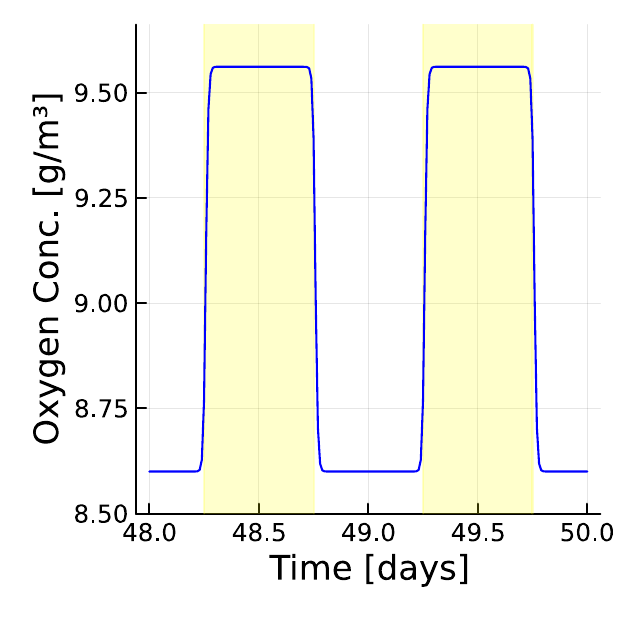}
    \end{subfigure}
    \quad
    \begin{subfigure}[b]{0.25\textwidth}
        \centering
        \includegraphics[width=\textwidth]{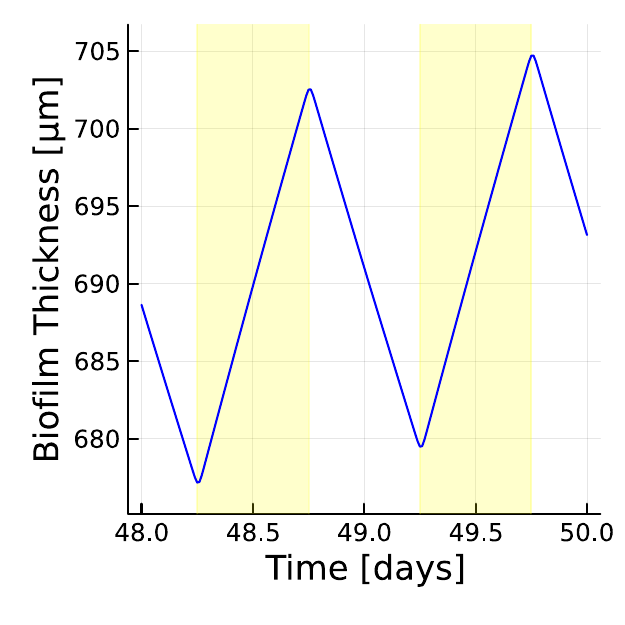}
    \end{subfigure}
    \caption{Phototroph and oxygen concentrations and biofilm thickness versus last two days of Case 5 simulation. Yellow regions indicate times when the light is on.}
    \label{fig:Case5_2days}
\end{figure}

Due to the high-frequency changes in the results, this example can be post-processed to observe additional features in the solution.  To plot just the last two days of the simulation, the \lstinline{biofilm_analyze()} function can be used again, but now with multiple times, \ie,
\begin{lstlisting}
# Times to analyze solution 
tout = 48.0:0.01:50.0; 
# Get solution at certain times
Xtout,Stout,Lfout = biofilm_analyze(sol,p,tout)
# Plot tank particulate concentration versus time 
plot(tout,Xtout')
\end{lstlisting}
which produces a simpler version of the left plot in Fig.~\ref{fig:Case5_2days}. The complete code to produce the three plots in Fig.~\ref{fig:Case5_2days} is available in \ref{sec:Case5_postprocessing}.

With these results, it is clear how the quantities vary with time.  When the light is on the phototroph (particulate) concentration increases and then decreases when the light is off.  The oxygen concentration is highest in the tank when the light is on and oxygen is generated by the growing phototroph.  When the light is off, the oxygen concentration decreases to 8.6 which matches the specified influent substrate concentration \lstinline{Sin}.  The biofilm thickness increases when the light is on due to the phototroph growth in the biofilm and then decreases when the light is off.

\section{Conclusions}
A modern implementation of a one-dimensional biofilm model is presented.  The software, known as Biofilm.jl, solves for the temporal dynamics of particulates (biomass) and substrates which are effected by biological phenomena including growth, metabolism, and competition as well as physical processes such as diffusion, convection, attachment and detachment.  The model is implemented in Julia~\cite{bezanson_julia_2017} and leverages the state-of-the-art differential equations solvers provided by DiffereentialEquations.jl library~\cite{rackauckas_differentialequationsjl_2017}. 

Example biofilms are provided to demonstrate the capabilities of Biofilm.jl.  The first example is of a single heterotroph growing in an environment with a nutrient. The second example demonstrates an acid stress response and has a biomass that consumes glucose and produces lactate which inhibits the biomass growth.  The third example demonstratess a single biomass species that grows and then dies.  The fourth example demonstrates a complex interplay of sufide-oxidizing bacteria (SOB), sulfate-reducing bacteria (SRB), and dead bacteria with the substrates oxygen, sulfate, and hydrogen sulfide.  Finally, a fifth example is of a phototroph in a diurnal cycle.  The examples highlight the variety of biofilms that Biofilm.jl can simulate as well as provide templates for postprocessing simulation results.

Biofilm.jl is provided with an open-source license and is hosted on github.  The repository contains detailed user manual with installation, execution, and postprocessing instructions as well as an overview of the governing equations.  Coding and Julia best practices, including documentation, unit tests, and plot recipes, are implemented.

\appendix

\section{Code Unit Tests}
\label{sec:unitTests}
Tests have been implemented to ensure the code is running correctly. The tests consist of 1) a series of unit tests that check specific parts of the code and compare the results to analytic solutions and 2) examples that are compared with previously computed answers to ensure a code modification does not introduce a bug.  Below are details on the unit tests and the analytic solutions. 

The tests can be run by cloning the GitHub repository, activating the package, entering the package manager by typing "]" at the Julia prompt, and then running \lstinline{test}.  In addition to manual tests, all the tests are executed automatically each time a new commit is added to the GitHub repository.

\subsection{Zero $L_L$}
Having the code behave robustly when the boundary layer within the tank is set to zero is a nice code feature.  This is achieved through Eq.~\ref{eq:Stop} which avoids dividing by zero.  The unit test \lstinline{test_zeroLL} runs a simulation with \lstinline{LL = 0.0} and checks that the substrate concentration at the top of the biofilm \lstinline{S_top} equals the substrate concentration at the topmost grid cell within the biofilm.  The test shows the condition is met to machine precision.

\subsection{Flux matching}
At the top of the biofilm the flux of substrate through the tank boundary layer matches the flux through the top of the biofilm to the first grid point, \ie, Eq.~\ref{eq:fluxMatch}.  This test ensures that the calculation of \lstinline{S_top} is correct and the fluxes do match computing \lstinline{S_top} then computing the two fluxes and ensuring they are equal. The test shows the condition is met to machine precision.

\subsection{Diffusion test}
The distribution of substrates within the biofilm is mainly controlled by diffusion and consumption due to the growth of particulates.  In this test, the steady-state distribution of substrate within the biofilm is compared with the analytic solution for a biofilm with one substrate and one particulate.  The particulate has a growth rate of $\mu = \mu_\mathrm{max} S /K_M$.  This biofilm, at steady-state, has an analytic solution of~\cite{hill_introduction_2014}
\begin{equation}
    S_{b:1}(z) = S_{t:1} \frac{ \cosh{ \phi z / Lf}}{\cosh{\phi}}
\end{equation}
where
\begin{equation*}
    \phi = \sqrt{\frac{\mu_\mathrm{max} \rho P_{b:1} L_f^2}{D_b K_m Y_{1,1}}}
\end{equation*}

The unit test consists of running a series of simulations until the biofilm reaches a steady state with different numbers of grid cells within the biofilm.  For each simulation, the error, which is the maximum difference between the analytic and computed solution, is computed.  Finally, the order of the method is computed by fitting a line to the error versus the number of grid points in log-log space.  It is expected the methods are second-order accurate and that the test shows an order of 1.9 for the grids tested. 

\subsection{Steady-State with Large Diffusivity}
If the diffusion coefficients $D_b$ and $D_{t}$ are considered to be very large, a steady-state solution to the biofilm model can be found since the substrate concentration can be assumed to be uniform within the biofilm and also equal to the tank concentration.  Note that in the following derivation of the analytic solution, subscripts on the substrate and particulate have been dropped as there is only one of each. 

The steady-state solution can be found by setting the time derivatives equal to zero in the differential equations.  For example, the 
thickness of the biofilm is described by Eq.~\ref{eq:dLfdt} and at steady-state this reduces to 
\begin{equation*}
    v(z_{N_z+1}) = v_\mathrm{det},
\end{equation*}
which means that the vertical velocity due to growth in the biofilm (LHS) matches the detachment velocity (RHS).  The LHS can be computed using Eq.~\ref{eq:growthvelocity_discrete} which when simplified, by setting $\mathrm{src}_X = 0$ and letting $S_b = S$ where $S$ is the uniform substrate concentration in the biofilm and tank, becomes $v(z_{N_z}) = \mu(S) L_f$.  The RHS can be evaluated with the definition of the detachment velocity in Eq.~\ref{eq:vdet}.  Combining these simplifications and solving for $L_f$ leads to 
\begin{equation}
    \label{eq:steady_Lf}
    L_f = \frac{\mu(S)}{K_\mathrm{det}}
\end{equation}
which provides the steady-state biofilm thickness. 

The particulate concentration in the tank is described by Eq.~\ref{eq:dXtdt}, which can be simplified by assuming steady-state, no source term, and using $v_\mathrm{det} = K_\mathrm{det} L_f^2 = \mu(S) L_f$ (Eq.~\ref{eq:steady_Lf}) leads to 
\begin{equation} 
    \label{eq:steady_Xt}
    0 = \mu(S) X_t - \frac{Q X_t}{V} + \frac{\mu(S) L_f A X_{b:N_z}}{V}.
\end{equation}

The steady-state substrate concentration can be simplified by taking Eq.~\ref{eq:dStdt} and letting the time derivative equal zero and setting the source term to zero.  Furthermore, the flux of substrate into the biofilm can be found by assuming $S_b=S$ and then integrating Eq.~\ref{eq:dSbdt} over the biofilm and using the divergence theorem which leads to 
\begin{equation*}
    S^\mathrm{flux}_{t} = \frac{\mu(S) X_b}{Y_{1,1}} L_f
\end{equation*}
and states that the flux of substrate into the biofilm is equal to the consumption used by the growth of particulates within the biofilm. 
All together these simplifications change Eq.~\ref{eq:dStdt} into 
\begin{equation*}
    0 = -\frac{\mu(S) X_t}{Y_{1,1}} + \frac{Q S_\mathrm{in}}{V} - \frac{Q S}{V} - \frac{\mu(S) X_b L_f A}{Y_{1,1} V}
\end{equation*}
which can be rearranged to provide an expression for $X_t(S)$, \ie,
\begin{equation}
    \label{eq:steady_St}
    X_t =  \frac{Y_{1,1}Q}{\mu(S) V}(S_\mathrm{in} - S)
        -  \frac{X_b L_f A}{V}
\end{equation}

To solve for the steady-state solution, an initial guess for the substrate concentration $S$ was made.  With this guess, the biofilm thickness $L_f$ can be computed with Eq.~\ref{eq:steady_Lf}, tank particulate concentration with Eq.~\ref{eq:steady_St}, and then the residual of Eq.~\ref{eq:steady_Xt} is evaluated.  Based on this residual, the guess for $S$ is updated and the procedure is repeated until the residual becomes less than a tolerance. 
Running this test shows that the simulated value of the substrate concentration is found to be within 0.004\% of the analytic value. 

\subsection{Time Integration}
This test ensures the time integration is implemented correctly by checking the substrate concentration in the tank with an inflow and outflow.  There is a particulate in the simulation, but the growth rate is set to zero so that it does not consume (or produce) substrate.  The diffusion coefficients are also set to very small numbers to limit the impacts of substrate entering or leaving the biofilm.  With these assumptions, Eq.~\ref{eq:dStdt} simplifies to 
\begin{equation}
\frac{d S_{t}}{dt} =  \frac{Q}{V} \left(S_\mathrm{in} - S_{t}\right)
\end{equation}
which has a solution
\begin{equation*}
    S_t(t) = S_\mathrm{in} + (S_t^0 - S_\mathrm{in}) e^{-\frac{Q}{V} t}
\end{equation*}
where $S_t^0$ is the initial condition.
Comparing the computed solution to the analytic solution shows that the maximum relative error in the solution is roughly 1e-6. 

\section{Case 5 Postprocessing Code}\label{sec:Case5_postprocessing}
This code was used to produce the plots in Figs.~\ref{fig:Case5_lighton} and~\ref{fig:Case5_2days} in Section~\ref{sec:Case5}.  The code can be executed after the simulation for Case 5 is run.
\begin{lstlisting}
# Biofilm quantities 
tout = 49.5 # Time when light is on
biofilm_analyze(sol,p,tout,makePlot=true,plotSize=(900,325))
# Could also make plot by directly calling recipe 
# biofilm_plot_film(sol([0,tout]),p,size=(900,325))
savefig("Case5_lighton.pdf")

# Times to analyze solution 
tout = 48.0:0.01:50.0; 
# Get solution at certain times
Xtout,Stout,Lfout = biofilm_analyze(sol,p,tout)

# Function to plot when the light is on
rectangle(w, h, x, y) = Shape(x .+ [0,w,w,0], y .+ [0,0,h,h])
function plot_light()
    plot(rectangle(0.5,1000.0,48.25,0.0), 
        opacity=.2, 
        fillcolor=:yellow,
        linecolor=:yellow,
        )
    plot!(rectangle(0.5,1000.0,49.25,0.0), 
        opacity=.2, 
        fillcolor=:yellow,
        linecolor=:yellow,
        )
end

# Plot tank particulate concentration versus time 
plot_light()
plot!(tout,Xtout',
    linecolor=:blue,
    xlabel=("Time [days]"),
    ylabel=("Phototroph Conc. [g/m3]"),
    legend=false,
    size=(300,300),
    ylims=(minimum(Xtout)-0.001,maximum(Xtout)+0.001),
    )
savefig("Case5_Xt.pdf")

# Plot tank substrate concentration versus time 
plot_light()
plot!(tout,Stout',
    linecolor=:blue,
    xlabel=("Time [days]"),
    ylabel=("Oxygen Conc. [g/m3]"),
    legend=false,
    size=(300,300),
    ylims=(minimum(Stout)-0.1,maximum(Stout)+0.1),
    )
savefig("Case5_St.pdf")

# Plot biofilm thickness versus time 
plot_light()
plot!(tout,1e6.*Lfout,
    linecolor=:blue,
    xlabel=("Time [days]"),
    ylabel=("Biofilm Thickness [um]"),
    legend=false,
    size=(300,300),
    ylims=(minimum(1e6*Lfout)-2,maximum(1e6*Lfout)+2),
    )
savefig("Case5_Lf.pdf")
\end{lstlisting}



\bibliographystyle{elsarticle-num}
\bibliography{references.bib}







\end{document}